\pdfoutput=1
\documentclass[12pt,fleqn]{article}
\usepackage{amsmath}
\usepackage{amssymb}
\usepackage{a4}
\usepackage[bf]{caption}
\usepackage{graphicx}

\begin{document}
\begin{center}
\vspace{1cm}
{\LARGE \textbf{Spin states of the first four holes in a silicon nanowire quantum dot}}\\
\vspace{1cm}
\textsl{{\large Floris A. Zwanenburg, Cathalijn E. W. M. van Rijmenam, \\ Leo P. Kouwenhoven$^\ast$}}\\
Kavli Institute of Nanoscience, Delft University of Technology, P.O. Box 5046, 2600 GA Delft, The Netherlands\\
\vspace{1cm}
\textsl{{\large Ying Fang, Charles M. Lieber}}\\
Department of Chemistry and Chemical Biology, Harvard University, Cambridge MA 02138, Unites States
\end{center}

\vspace{1cm}
\noindent $^\ast$ corresponding author: l.p.kouwenhoven@tudelft.nl\\

\vspace{1cm}

\noindent ABSTRACT: We report measurements on a silicon nanowire quantum dot with a clarity that allows for a complete understanding of the spin states of the first four holes. First, we show control of the hole number down to one. Detailed measurements at perpendicular magnetic fields reveal the Zeeman splitting of a single hole in silicon. We are able to determine the ground-state spin configuration for one to four holes occupying the quantum dot and find a spin filling with alternating spin-down and spin-up holes, which is confirmed by magnetospectroscopy up to 9T. Additionally, a so far inexplicable feature in single-charge quantum dots in many materials systems is analyzed in detail. We observe excitations of the zero-hole ground-state energy of the quantum dot, which cannot correspond to electronic or Zeeman states. We show that the most likely explanation is acoustic phonon emission to a cavity between the two contacts to the nanowire.
\clearpage

Long spin lifetimes are crucial for applications such as spintronics \cite{wolf} and even more so for quantum computation with single spins. The proposal to use single spins as quantum bits \cite{loss, kane} exploits an optimal combination of the spin and charge degree of freedom \cite{hansonrmp}. The potential of this spin qubit is underlined by the recent demonstration of coherent control of one \cite{koppens} and two \cite{petta} spin states in quantum dots in GaAs/AlGaAs heterostructures. Most experiments have focused on quantum dots formed in III–-V semiconductors; however, electron spin coherence in those materials is limited by hyperfine interactions with nuclear spins and spin-orbit coupling. Group IV materials are believed to have long spin lifetimes because of weak spin-–orbit interactions and the predominance of spin-zero nuclei. This prospect has stimulated significant experimental effort to isolate single charges in carbon nanotubes \cite{jarillo, mceuen}, Si FinFETs \cite{rogge} and Si nanowires \cite{zhong}. The recent observation of spin blockade in Si/SiGe heterostructures is argued to confirm the predicted long-lived spin states \cite{eriksson}. Here we identify the spin states of single charges in silicon quantum dots by means of low-temperature electronic transport experiments, for the first time to the level of individual spin states.\\

We have measured 30 Si nanowire quantum dots with pronounced excited states. In the measurements presented here, both a backgate and a side gate allow control of the number of charges down to a single hole in the dot. We observe the Zeeman energy of the first two holes at magnetic fields ranging from 0 to 9 T, from which we deduce a $g$-factor close to the Si bulk value. Magnetospectroscopy of the first four holes allows determination of the successive spins that are added to an empty dot and reveals a spin filling with alternating spin-down and spin-up holes. The isolation and identification of a single spin in silicon demonstrated here constitutes an important step towards spintronic applications in a material with a long spin coherence time. Additionally, we have identified many excited-state lines which cannot be attributed to electronic states of the quantum dot itself. The discrete energy level spectrum is found to be independent of magnetic field and is most probably caused by spontaneous emission to a phonon cavity. \\

Single-crystal $p$-type Si nanowires are prepared by a gold nanocluster mediated vapor-liquid-solid process \cite{wagner}, using silane and diborane as precursor gases with an atomic feed-in ratio of Si:B = 4000:1 \cite{zhong}. The typical diameter is 7-12 nm, comprising a Si core of 3-8 nm and a native oxide of $\sim$2 nm. After growth we deposit the nanowires on a highly doped silicon substrate capped with a dry thermal oxide. Predeposited markers allow locating individual nanowires with an SEM and defining contacts by means of electron-beam lithography. The samples are treated with buffered hydrofluoric acid for 5 s prior to metal deposition to etch off the native SiO$_2$. We then evaporate 60-100 nm thick Ni contacts, leaving a Si channel of typically 300 nm uncovered. After metal lift-off the samples are annealed in sequential steps of 20-30 seconds at 380$\,^{\circ}{\rm C}$ and 410$\,^{\circ}{\rm C}$, to induce radial and longitudinal diffusion of Ni into the Si nanowire. From both Ni contacts a NiSi segment is formed in the nanowires with lengths of 100-150 nm, depending on diameter, temperature and time. In Figure 1a a schematic of a resulting NiSi--Si--NiSi nanowire is shown. The remaining Si section is connected to the lithographically defined Ni contacts by two NiSi leads. Scanning electron micrographs reveal silicide segments as bright regions which sandwich a darker section of Si, see e.g. Figure 1b or \cite{hu, weber}. These devices have room-temperature resistances varying from 100 k$\Omega$ to 5 M$\Omega$. They are cooled down to cryogenic temperatures with a pumped $^4$He-cryostat or a dilution refrigerator. All data in this paper have been taken at base temperature of 20 mK. Some nanowires are fully transformed into NiSi having room temperature resistances of 1-5 k$\Omega$, corresponding to $\rho\sim10$ $\mu\Omega$cm, consistent with values found in NiSi nanowires \cite{wu} and large single crystals \cite{meyer}.

Electrical characterization is carried out by measuring the current from drain to ground while sweeping the bias voltage at the source, $V_{SD}$, and stepping the backgate voltage, $V_{BG}$. The resulting differential conductance, $dI/dV_{SD}$, versus $V_{SD}$ and $V_{BG}$ shows a set of diamond-shaped regions, in which the current is zero due to Coulomb blockade \cite{21}. Typical measurements are presented in Supplementary Figure 1. Inside a Coulomb diamond the number of charges, $N$, on the dot is fixed. The diamond edges mark the onset of a finite current when the ground state of the $N^{\textrm{th}}$ hole, GS($N$), becomes available for transport and the number of holes starts to alternate between $N$ and $N-1$. Outside the diamonds many lines run parallel to the edges, indicating a change in conductance which is caused by the availability of extra channels for transport. Note that lines ending on the $N^{\textrm{th}}$ diamond are attributed to the excited states of the $N^{\textrm{th}}$ hole, ES($N$) \cite{hansonrmp}. The fact that excited states are visible is a direct consequence of the small size of the quantum dots and therefore large level spacing.\\

The relatively high electron and hole effective mass in silicon results in a relatively small energy level spacing. As a consequence, the observation of quantum states in silicon nanowires requires very short channel lengths and there are only few reports of pronounced excited states in Si devices \cite{rogge, zhong, eriksson, angus}. We have fabricated NiSi--Si--NiSi nanowires with a technique that allows the formation of dots shorter than 30 nm for which the stability diagrams of two representative devices are shown in the Supporting Information. The (near) degeneracy of heavy and light hole bands in silicon can be observed through different capacitances as a result of different charge distributions of the corresponding orbitals, see Supplementary Figure 1b and 1c. These results show that we can reproducibly fabricate Si nanowire quantum dots with sizes ranging from 10 to 100 nm. \\

In order to gain additional tuneability we have also fabricated devices with an extra side gate (inset of Figure 2a). In Figure 2a we plot the current versus backgate voltage, $V_{BG}$, and side gate voltage, $V_{SG}$. Each time a hole is added to the quantum dot, a current peak appears as a diagonal line, a typical signature of a single quantum dot \cite{vanderwiel}. The bending of two adjacent Coulomb peak lines towards or away from each other means that the addition energy changes. Apparently the shape of the confinement potential is modified differently by $V_{BG}$ and $V_{SG}$ because of their global (backgate) or more local electric field (side gate). As a result the potential well is not a perfect parabola as sketched in Figure 1a. Some peaks become switchy over a certain gate voltage range, see e.g. the bistable behavior for values of $V_{BG}$ between -18 and -12 V, but the device is very stable in the greater part of Figure 2a. Figure 2b shows a stability diagram, where the backgate voltage is used to vary the number of holes from 12 to 2.

In Figure 2c we reduce the number of charges down to one with the side gate and the backgate. The last diamond opens completely up to $\pm$ 200 mV bias in both diagrams. We have measured no current up to $V_{BG}$=50 V, which means that the last diamond does not close and we have indeed observed the last hole (Figure 3 provides additional proof). The excited state of the first hole lies about 80 meV above the ground state. This is ideal for spin qubits, where individual spin states need to be addressed without interference of higher orbitals. The results in Figure 2 demonstrate a high degree of device stability and tunability, providing control of the number of holes down to a single-hole quantum dot. \\

The observation of the last hole allows us to identify the first four spin states on the quantum dot. We zoom in on the $N=0\leftrightarrow1$ and the $N=1\leftrightarrow2$ transition at $B$ = 0, 1, 2, 4 and 8 T applied perpendicular to the substrate (Figure 3a). Both transitions exhibit a set of excited states parallel to the diamond edges at 0 T, spaced by roughly 100 $\mu$eV. They cannot be attributed to electronic states, since these are higher in energy than 10 meV, as shown in Figure 2c. We discuss those below, but first focus on the Zeeman energy $E_Z = |g|\mu_BB$, where $g$ is the $g$-factor and $\mu_B$ is the Bohr magneton. A finite magnetic field splits the spin-degenerate ground-state transport line into a spin-up and spin-down line separated by $E_Z$. In our measurement the Zeeman-split spin state can be distinguished from the other excited states by its magnetic field dependence. Figure 3b shows the Zeeman splitting extracted from measurements as in Figure 3a at magnetic fields up to 9 T. Linear fits yield measured $g$-factors of $\sim$2.3. We have also measured the Zeeman energy of a single-hole in another device, see Supplementary Figure 2.

At the $N=0\leftrightarrow1$ ($N=1\leftrightarrow2$) transition, the Zeeman-split spin state appears only at positive (negative) voltage bias, which we can relate to asymmetric source and drain tunnel barriers \cite{cobdenspin}, as illustrated in Figure 3c. In contrast to the edges of the $(N=0)$ and $(N=2)$-diamonds, the $(N=1)$-diamond edges do not show Zeeman splitting.  This means that in the two-hole ground-state both holes are in the lowest orbital with opposite spins \cite{cobdenshell, laurens}, i.e. a singlet state. Assuming the $g$-factor is positive, as in bulk Si, the first hole ground-state is spin-down, $\mid\downarrow\rangle$.  When the second hole is added to the lowest orbital it is spin-up, $\mid\uparrow\rangle$. At the $N=2\leftrightarrow3$ and the $N=3\leftrightarrow4$ transitions the even-$N$ diamond edges split (see Supplementary Figure 3), which implies that the third and fourth hole are respectively spin-down and spin-up \cite{laurens}.\\

In addition to the excited-state spectroscopy of Figure 3, we have performed magnetospectroscopy on the ground states of the first four holes to confirm the spin filling of the quantum dot. Figure 4a shows the evolution of the first four Coulomb peaks as the magnetic field is stepped from 0 to 9 T. The current peaks move towards or away from each other, depending on the spin direction of the additional hole. In Figure 4b we plot the distance between consecutive peaks, $\Delta V_{SG}$, for 1, 2 and 3 holes. The alternating direction of the Coulomb peak evolution shows that spin-down and spin-up holes alternately enter the dot \cite{folk}.\\

Up to now we have identified excited-state lines that are caused by orbital or spin states of the quantum dot. This leaves the additional excitations spaced by roughly 100 $\mu$eV in Figure 3a. Lines ending on the $N=0$ region correspond to excitations of the zero-hole system. The zoom on the $N=0\leftrightarrow1$ transition in Figure 5a displays equally spaced lines at energies $E_n$, to which we have assigned a number $n$. Line cuts in Figure 5b and 5c reveal a set of small peaks in the current-voltage characteristics. The difference in energy between the $n^{th}$ line and the ground-state transport line, $E_n-E_0$, is plotted as a function of $n$ in the upper left inset (for $E_n$ we take the center of each line). Analogously the energy level spacings for different magnetic fields are plotted in the lower right inset. Clearly the lines are not affected by a magnetic field. The same analysis of the $N=1\leftrightarrow2$ transition gives similar results, see Figure 5c. The $N=2\leftrightarrow3$ and the $N=3\leftrightarrow4$ transition also show discrete spectra independent of magnetic field, see Supplementary Figure 3.

The additional lines do \textsl{not} originate from electronic states of the Si quantum dot itself, and are therefore necessarily due to interactions with the environment. The origin of the energy spectrum lies in inelastic tunneling processes via discrete energy levels separated by $n\Delta E$ from the elastic ground-state transport level. In this scenario, additional tunneling processes exist where packets of energy, $\Delta E$, are emitted into the environment as illustrated in the diagrams in Figure 5d. The most likely explanation is spontaneous energy emission to a phonon cavity of $\sim$100 nm long (see figure caption). This lengthscale corresponds well to the total length of NiSi--Si--NiSi nanowire of 250 nm. The cavity edges are then situated at the transition from the amorphous Ni contacts to the crystalline NiSi, where the cross-sectional area changes stepwise by more than three orders of magnitude.

Absorption of phonons from the environment can, however, be ruled out for two reasons: (i) The lines stop at the diamond edges, whereas phonon absorption would also be visible in the Coulomb blockaded regions. (ii) The height of the small current peaks due to inelastic tunneling in Figure 5b and 5c is $\sim$1 pA, corresponding to one charge leaving the dot every 160 ns. It takes a phonon $\sim$40 ps to go up and down a 100 nm cavity, based on a speed of sound of 5000 ms$^{-1}$. Absorption of previously emitted phonons thus requires a cavity with a Q-factor of $\sim$4000, which is unrealistically high in our system.

An energy emission of $n\Delta E$ can be interpreted by two possible scenarios: (i) one phonon with energy $n\Delta E$ is sent out. Each line corresponds to one phonon mode of energy $n\Delta E$, hence the modes are equidistant in energy as in a harmonic oscillator potential. (ii) $n$ phonons with energy $\Delta E$ are emitted. Since we observe emission even above 10$\Delta E$ (see Figure 5), the electron-phonon coupling must be very strong in the latter scenario \cite{flensberg}. According to the Frank-Condon model one would expect current steps, as seen in C$_{60}$-molecules \cite{park}, instead of the observed peaks. Both peaks and steps have been reported in suspended carbon nanotubes \cite{sapmaz}. The precise nature of the phonon-assisted tunneling processes requires a detailed explanation and thus provides an interesting challenge for theory.

\noindent
\textbf{Acknowledgements}\\
\noindent
We thank T. Balder, R. Hanson, A. A. van Loon, K. C. Nowack, R. N. Schouten, G. A. Steele and I. T. Vink for help and discussions. Supported by the Dutch Organization for Fundamental Research on Matter (FOM), the Netherlands Organization for Scientific Research (NWO) and NanoNed, a national nanotechnology program coordinated by the Dutch Ministry of Economic Affairs. C.M.L. acknowledges Samsung Electronics and a contract from MITRE Corporation for support of this work. \\

\clearpage
\noindent
{\Large \textbf{Figure captions}}\\

\noindent \textbf{Figure 1: Small silicon quantum dots in NiSi--Si--NiSi nanowires.} (\textbf{a}) Schematic of a Si nanowire quantum dot with NiSi leads on an oxidized Si substrate. The NiSi leads are fabricated by thermally induced Ni diffusion from the two Ni contacts into the Si nanowire. Lower panel shows an energy diagram of the corresponding Schottky tunnel barriers that define the quantum dot and the resulting discrete energy levels in the dot. Occupied (empty) hole states are indicated in red (blue).  (\textbf{b}) SEM of an actual device, where the NiSi shows up brighter than the Si.\\

\noindent \textbf{Figure 2: Observation of the last hole.} (\textbf{a}) Current in color scale versus side gate voltage, $V_{SG}$, and backgate voltage, $V_{BG}$, at a bias of 2 mV. Diagonal lines correspond to transitions from $N$ to $N+1$ holes, indicated in white digits. The slopes give roughly $C_{SG}=1.3C_{BG}$, with $C_{BG}\sim 0.07aF$, based on diamond 9 in Figure 2b. The peak of the last hole (the $N=0 \leftrightarrow 1$ transition) is about 5 pA high and as a result barely visible in this color scale. Inset: SEM picture of the device with two Ni contacts and a SiO$_2$/Cr/Au side gate (SG). The side gate is about 40 nm away from the nanowire and the distance between the lithographically defined Ni contacts is 250 nm. The nanowire broke after the measurements. (\textbf{b}) Stability diagram of the same device, showing $dI/dV_{SD}$ in color scale versus $V_{SD}$ and $V_{BG}$ ($V_{SG}=0$ V, i.e. along the white dashed line in (\textbf{a})). Based on the backgate capacitance at high hole numbers we estimate the Si dot length to be about 12 nm (see Supporting Information), becoming even smaller as holes leave the dot. (\textbf{c}) Either gate can be used to control the number of holes down to zero. In both cases the last diamond opens completely, a strong indication of an empty quantum dot.\\

\noindent \textbf{Figure 3: Zeeman energy of the first two holes.} (\textbf{a}) Zoom on the 0$\leftrightarrow$1 and the 1$\leftrightarrow$2 transition at $B$ = 0, 1, 2, 4 and 8 T. The Zeeman line (blue arrows) moves away from the ground-state line (white arrows) with $|g|\mu_BB$, and appears at positive (negative) bias at the 0$\leftrightarrow$1 (1$\leftrightarrow$2) transition. (\textbf{b}) Zeeman energy versus magnetic field for both transitions. Black and red lines are linear fits corresponding to respectively $|g|=2.27\pm0.18$ and $|g|=2.26\pm0.23$. (\textbf{c}) Energy level diagrams explaining the influence of asymmetric barriers when the Zeeman-split state enters the bias window. Occupied (empty) hole states are indicated in red (blue). At the 0$\leftrightarrow$1 transition there is only an observable increase at positive bias. When the spin-excited state enters the bias window, the tunnel rate onto the dot increases from $\Gamma_{\downarrow}^{\textrm{in}}$ to $\Gamma_{\uparrow}^{\textrm{in}} + \Gamma_{\downarrow}^{\textrm{in}}$ (left diagram) because a hole with either spin-up or spin-down can enter. The tunnel rate to leave the dot $\Gamma^{\textrm{out}}$ does not change since only one hole can tunnel off (rightmost diagrams): $\Gamma^{\textrm{out}}$ = $\Gamma_{\uparrow}^{\textrm{out}}$ or $\Gamma_{\downarrow}^{\textrm{out}}$, depending on which spin entered the dot previously (assuming no spin relaxation). This means that the addition of an extra level in the bias window only increases the conductance noticeably if the holes tunnel into the dot via the barrier with the lowest tunnel rate. On the contrary, in the transport cycle of the 1$\leftrightarrow$2 transition $\Gamma^{\textrm{in}}$ does \textsl{not} change when an additional level enters the bias window, since only one spin species can enter the dot (leftmost diagrams). With two spins on the dot, this singlet state can change to either $\mid\downarrow\rangle$ or $\mid\uparrow\rangle$ when respectively a spin-up hole or a spin-down hole leaves the dot. Since both spin species can tunnel off, $\Gamma_{\textrm{out}}$ increases to $\Gamma_{\uparrow}^{\textrm{out}} + \Gamma_{\downarrow}^{\textrm{out}}$ when the second level becomes available for transport. This is reflected by the Zeeman-split line at negative bias, where holes tunnel \textsl{off} via the barrier with the lowest tunnel rate. \\

\noindent \textbf{Figure 4: Magnetospectroscopy of the first four holes.} (\textbf{a}) Magnetic field evolution of Coulomb peaks corresponding to the addition of hole 1 (upper panel, $V_{SD}=0.8$ mV) and hole 2, 3 and 4 (lower panel, $V_{SD}=0.6$ mV). The evolution alternates due to even-odd spin filling. (\textbf{b}) The peak-to-peak distance $\Delta V_{SG}$ versus magnetic field and linear fits for 1, 2 and 3 holes. An offset has been subtracted to reveal the evolution. Conversion of the slopes of the linear fits to the Zeeman energy via the gate coupling factor $\alpha$ yields $g$-factors of $1.72\pm0.29$, $1.77\pm0.43$ and $2.02\pm0.58$. The deviation from the numbers found in Figure 3 is caused by the fact that our diamond shapes are not perfect parallelograms. This makes a strict definition of $\alpha$ for a single hole number impossible. For this estimate we use $\alpha = \Delta V_{SD}/2\Delta V_{BG}$.\\

\noindent \textbf{Figure 5: Quantized energy emission to the environment.}
(\textbf{a}) Differential conductance in color scale versus $V_{SD}$ and $V_{SG}$ at the $N=0\leftrightarrow1$ transition at 9 T. Blue arrows and numbers point to lines of increased conductance. (\textbf{b}) Current-voltage characteristics taken as line cuts from (\textbf{a}) at $V_{SG}$ = 2288 mV (blue trace, offset by 1 pA, corresponding to dashed blue line in (\textbf{a})) and 2301 mV, (green trace, corresponding to dashed green line in (\textbf{a})). Upper left inset: energy difference between the $n^{\textrm{th}}$ line and the ground-state energy versus $n$ for negative bias at 9 T (red circles). Lower right inset: energy level spectrum at other magnetic fields. Black trace is a linear fit with a slope of 111 $\mu$eV. (\textbf{c}) Current-voltage characteristics taken as line cuts from Figure 3a at $V_{SG} = -3788$ mV (green trace) and $-3768$ mV (blue trace, offset by 5 pA). Upper left inset: energy difference between the $n^{th}$ line and the ground-state energy at 0 T versus $n$ for positive bias (black circles) and negative bias (red circles). The linear fits through the data points both have a slope of 126 $\mu$eV (black and red trace, not separately visible). Lower right inset: Energy level spectrum at finite magnetic fields. Black trace is a linear fit with a slope of 117 $\mu$eV. (\textbf{d}) The middle panel sketches the measured differential conductance of (\textbf{a}) in straight lines (we leave out the Zeeman-split excited state); dashed lines are energy levels that do not contribute significantly to the current. They do, however, appear at the $N=2\leftrightarrow3$ and the $N=3\leftrightarrow4$ transition, because the tunnel barriers are more symmetric, see Supporting Information. The adjacent diagrams show the possible tunnel processes when the ground-state transport level is aligned to source (blue symbols) or drain (green symbols). Black (red) arrows correspond to tunneling processes without (with) energy emission $\Delta E$. The tunnel rate into (out of) the dot increases for lines ending on the $N=1$ ($N=0$) region. Due to an asymmetry in tunnel barriers the inelastic tunnel processes only enhance the conductance if the holes tunnel inelastically in (out) through the barrier with the lowest tunnel rate. We stress that the electrochemical potential levels in the dot indicated by red dashed lines are \textsl{not} available states for holes residing on the dot: only the ground state can be occupied (black straight line). If the energy is emitted to a phonon cavity, its magnitude is determined by the phonon speed, $v$, and the length of the cavity, $L$, according to $\Delta E = hv/2L$. This allows us to estimate the order of magnitude of the cavity size required for an emission of $\Delta E\sim100$ $\mu$eV. Based on a speed of sound of $\sim$5000 ms$^{-1}$, the phonon cavity must be $\sim$104 nm long.

\clearpage
\noindent
{\Large \textbf{Figures}}\\
\begin{figure}[h]
\centering
\caption{ \ \ } \textbf{Small silicon quantum dots in NiSi--Si--NiSi nanowires} \\
\vspace{1cm}
\includegraphics[width=0.5\textwidth]{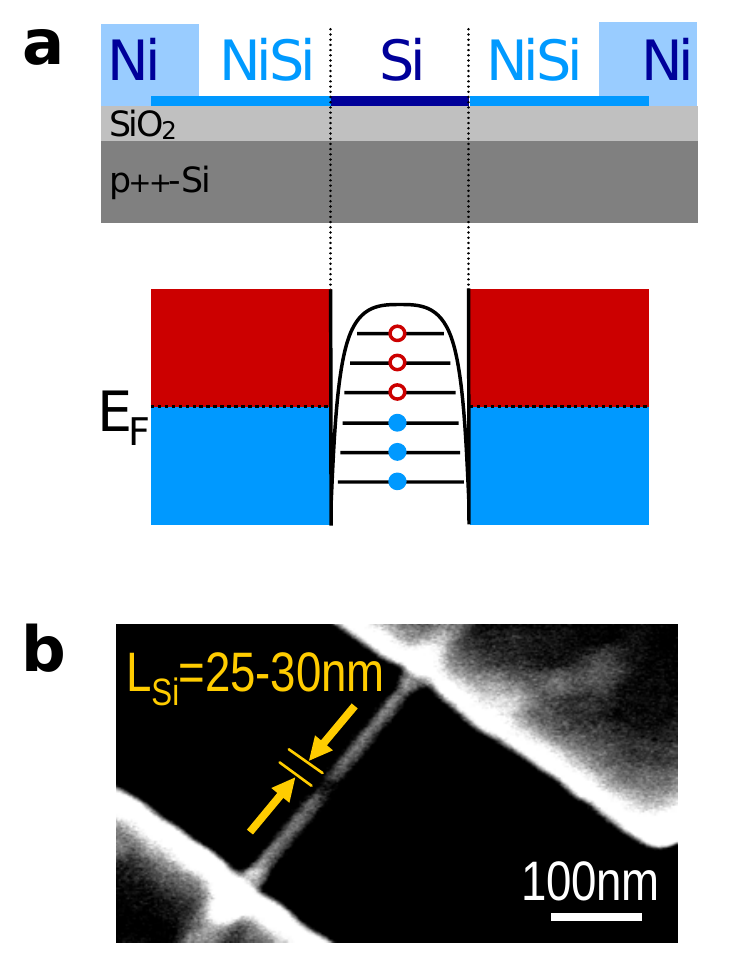}
\label{fhdfig1}
\end{figure}

\begin{figure}[t]
\centering
\caption{ \ \ } \textbf{Observation of the last hole} \\
\vspace{1cm}
\includegraphics[width=\textwidth]{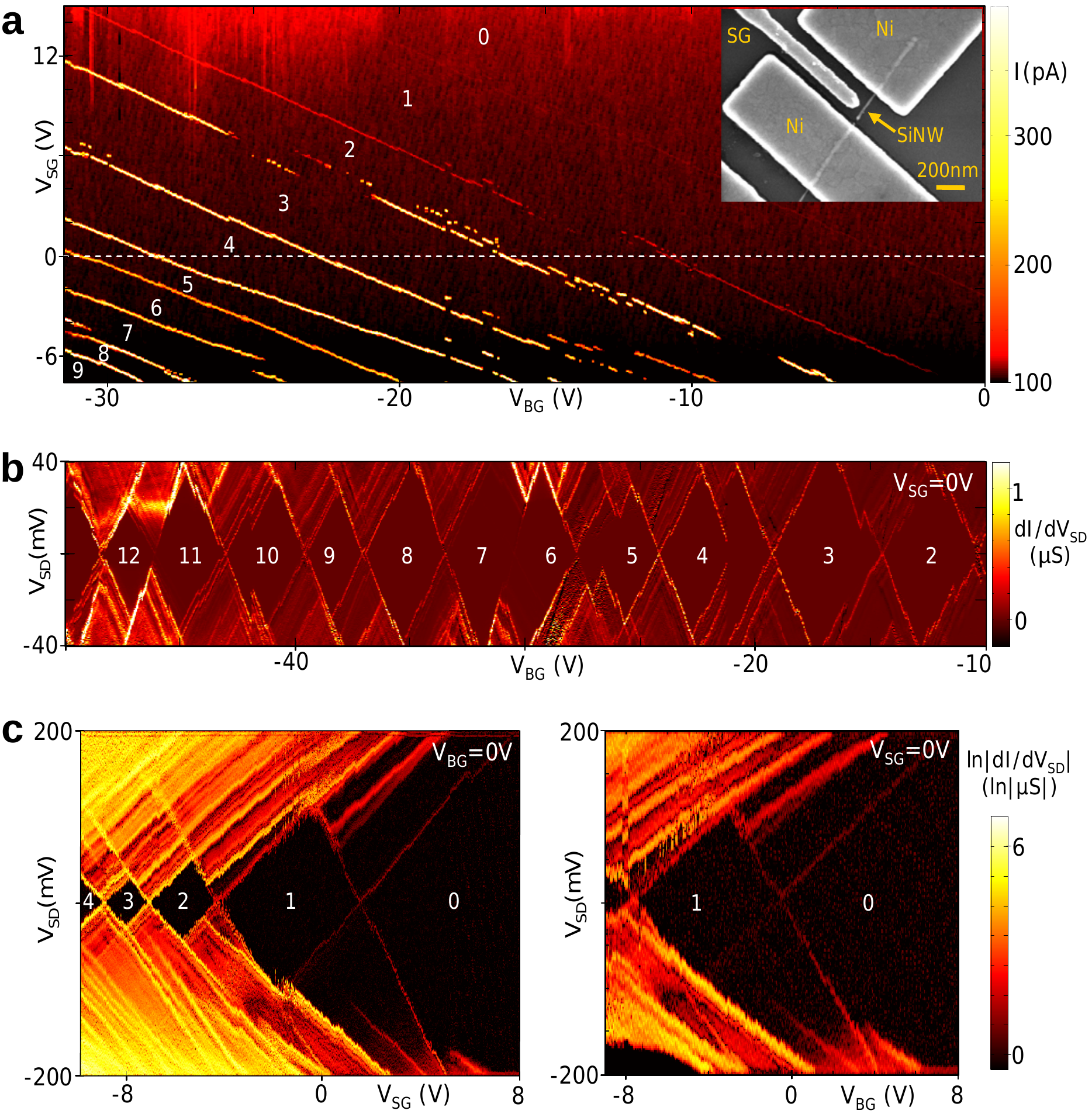}
\label{fhdfig2}
\end{figure}

\begin{figure}[t]
\centering
\caption{ \ \ } \textbf{Zeeman energy of the first two holes} \\
\vspace{1cm}
\includegraphics[width=\textwidth]{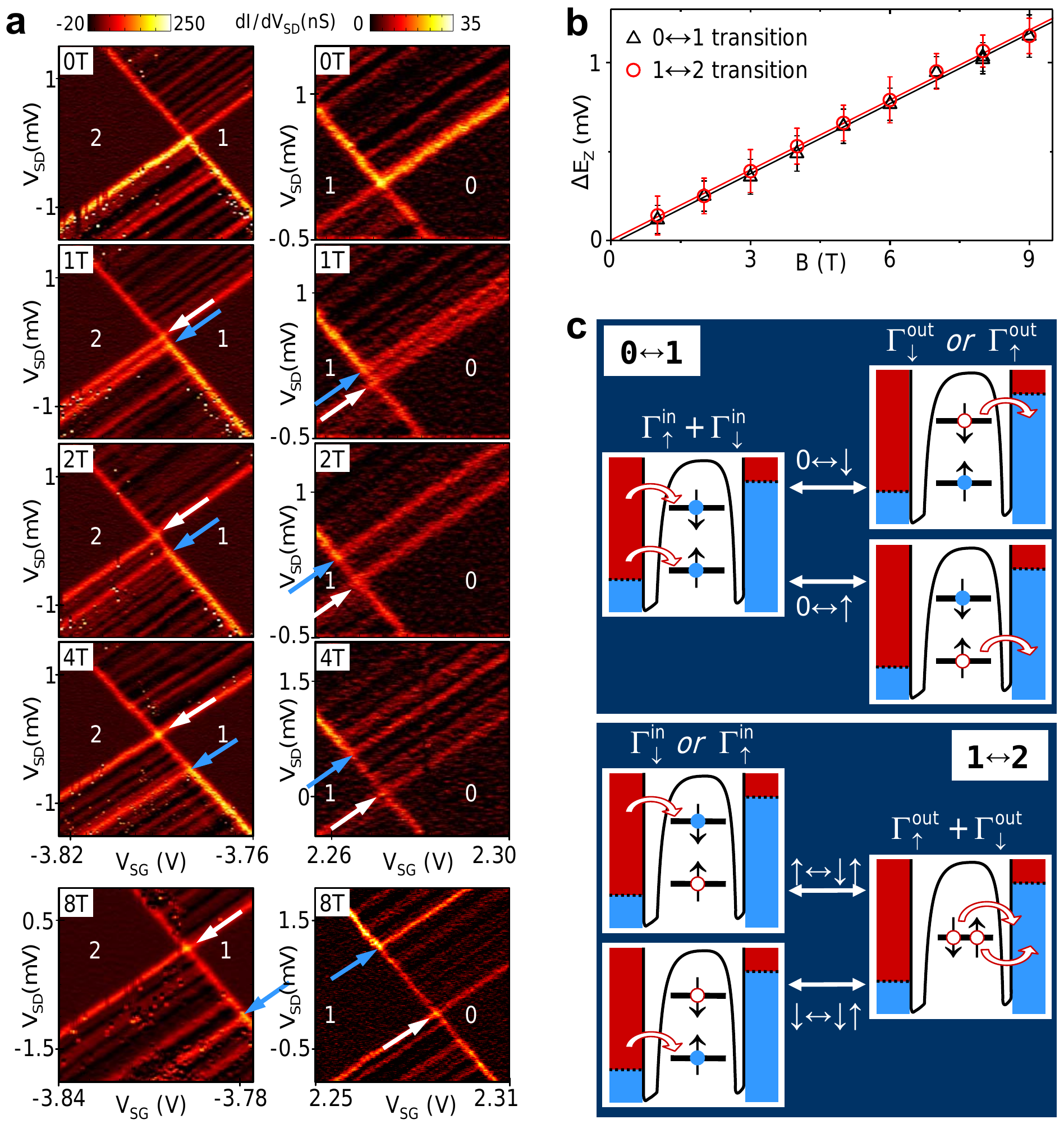}
\label{fhdfig3}
\end{figure}

\begin{figure}[t]
\centering
\caption{ \ \ } \textbf{Magnetospectroscopy of the first four holes} \\
\vspace{1cm}
\includegraphics[width=.8\textwidth]{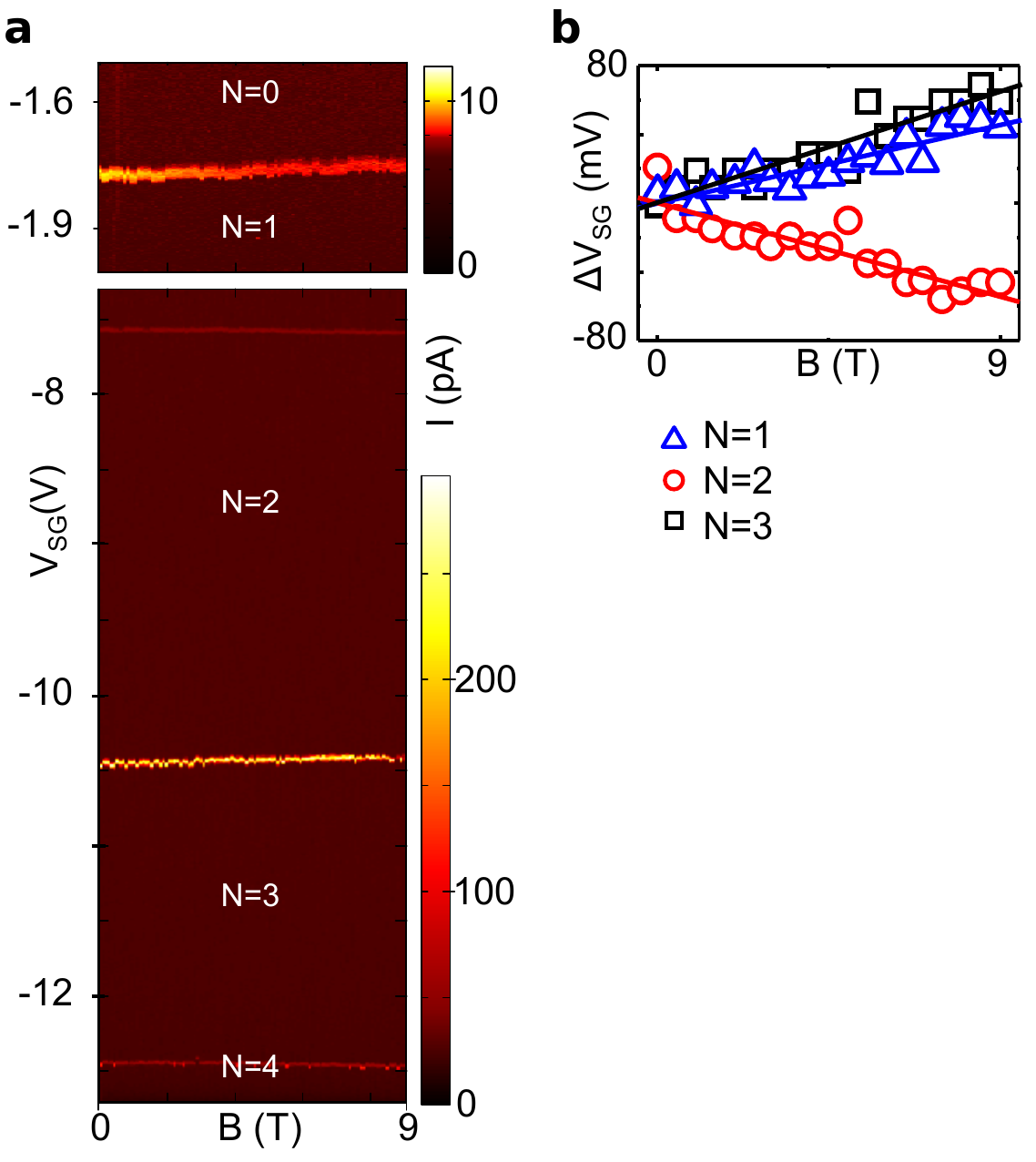}
\label{fhdfig4}
\end{figure}

\begin{figure}[t]
\centering
\caption{ \ \ } \textbf{Quantized energy emission to the environment} \\
\vspace{1cm}
\includegraphics[width=\textwidth]{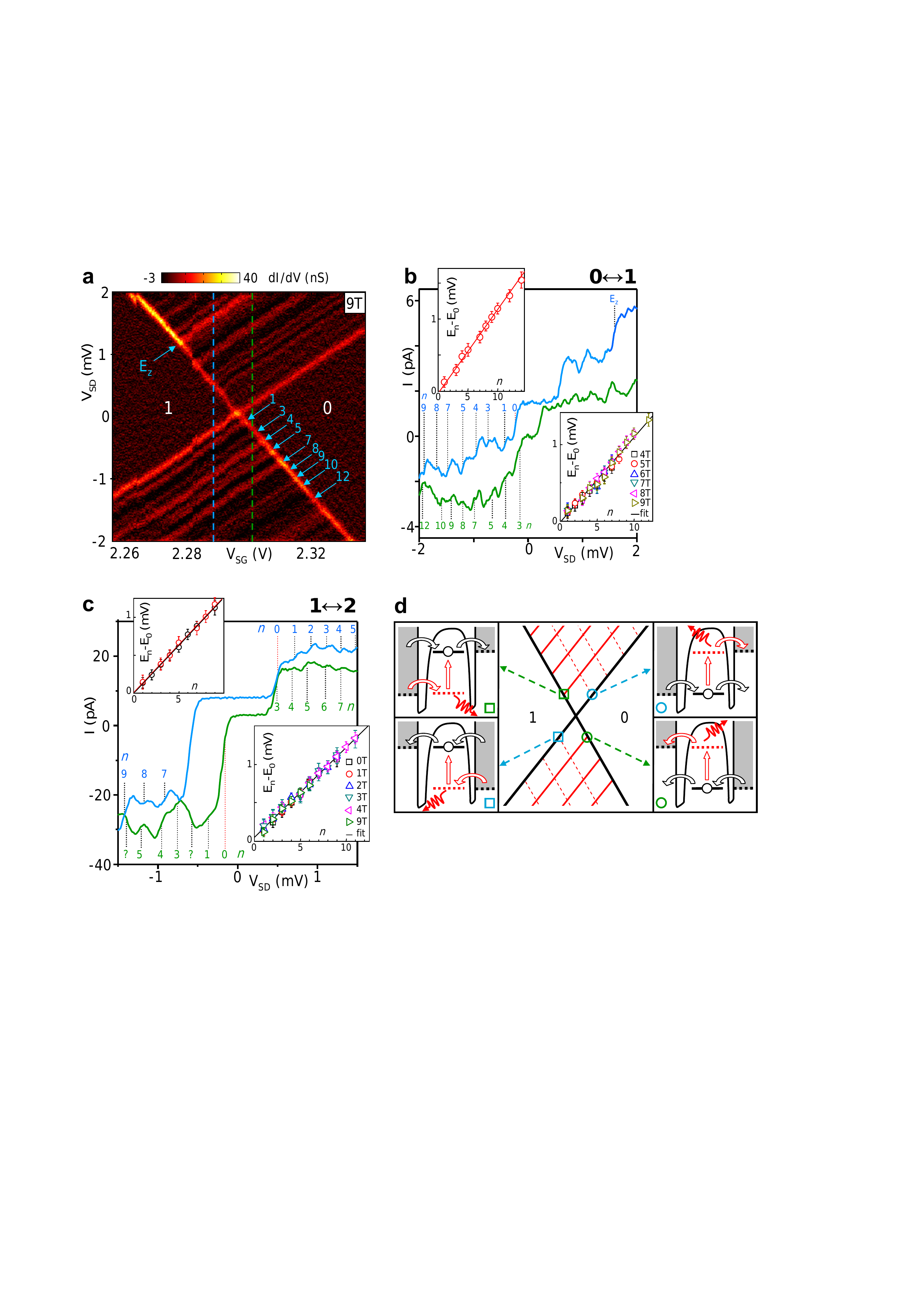}
\label{fhdfig5}
\end{figure}

\clearpage
\noindent
{\Large \textbf{Supporting information}}\\

\noindent
\textbf{Single quantum dots in Si nanowires}\\
\noindent
Electron-beam lithography is used to define Schottky tunnel contacts with separations down to 30-50 nm. The minimum width of the source and drain contacts is at least ten times the nanowire diameter. The small wire dimensions make effective gating difficult. In this geometry the distance to the backgate is greater than the dot length, and thus the electric field lines from the backgate are severely screened by the metallic leads. Instead, we have fabricated NiSi--Si--NiSi nanowire, where the leads and quantum dot have comparable diameters (see Figure 1), strongly reducing screening of the backgate compared to lithographically defined leads. In this case the backgate has a stronger capacitive coupling. An extra advantage is that this technique allows the formation of dots shorter than 30 nm.

The backgate-to-dot capacitance, $C_{BG}$, is derived as $C_{BG}=e/\Delta V_{BG}$, where $\Delta V_{BG}$ is the backgate voltage needed to add a single charge. A cylinder-on-plate model can be used to relate $C_{BG}$ to the Si dot length, $L$, see e.g. \cite{martel}. However, this model does not take into account screening of electric fields. To avoid this we have applied the Poisson equation to the geometry of Figure 1a to compute $C_{BG}(L)$, yielding e.g. $L$ = 28 nm for the device in Figure 1b. When the channel length measured in SEM images is compared to the length determined from the model we obtain an accuracy within 30\% in different devices. We conclude that the model gives a good estimate of the dot length and that the screening from contacts at NiSi--Si junctions is very small.\\

\noindent
\textbf{Quantum dots with larger hole numbers}\\
\noindent
The stability diagram in Supplementary Figure 1a displays a typical set of regular Coulomb diamonds that are very similar to other material systems. Supplementary Figure 1b, however, shows ten Coulomb diamonds of another device, which are kite-shaped: the slopes of two adjacent diamonds are not parallel, in contrast to the conventional parallelograms of Supplementary Figure 1a. The diamond edges have slopes of $-C_{BG}/C_S$ and $C_{BG}/(C-C_S)$, see ref \cite{hansonrmp}. Here $C=C_S+C_D+C_{BG}$ is the sum of all capacitances to the dot\footnote{We assume that no other gates have a significant capacitive coupling to the dot.} and $C_S$ ($C_D$) is the capacitance between dot and source (drain). Within this model a difference in slopes for two successive transitions is the result of different capacitances to the corresponding electronic orbitals.

Combining the above equations with the measured slopes gives source and drain capacitances between 1.7 and 2.2 aF for Supplementary Figure 1a. In Supplementary Figure 1b the spread is much wider and values vary from 2 to 10 aF, as shown by the calculated capacitances in Supplementary Figure 1c. In a quantum dot the specific shape of the orbital wave function determines its capacitive coupling to a metallic gate. $C_{BG}$ is roughly the same for different orbitals, because the backgate is relatively far away. On the other hand, since the source and drain are extremely close to the quantum dot and the shape of the wave function changes with each orbital, $C_S$ and $C_D$ can vary strongly. As can be seen in Supplementary Figure 1c, there are pairs of source and drain capacitances that have comparable values, indicating a symmetric orbital wave function. Conversely, there are pairs where $C_S$ is much greater than $C_D$, as a result of an orbital coupling which is much stronger to the source. We remark that the kites cannot be explained by multiple-dot behavior \cite{kubatkin} since all diamonds close at zero bias.

The slopes we observe for excited states are consistent with this picture: the excited-state line of the $M^{\textrm{th}}$ diamond is parallel to the ground-state line of the transition from $M$ to $M+1$ holes (indicated by blue arrows in Supplementary Figure 1b). In both cases the $(M+1)^{\textrm{th}}$ orbital is used for hole transport, resulting in the same capacitive coupling to source, drain and gate. However, the ground-state line of the transition from $M-1$ to $M$ holes has a different slope (green arrows), because it uses a different orbital. In the stability diagram of Supplementary Figure 1b we can find six pairs of excited-state lines parallel to the ground-state line of the next hole (grey arrows), with a slope different from the adjacent hole ground-state line (black arrows). In short, the kites Supplementary Figure 1b can be explained by different capacitances to different orbitals.

As far as we know kites have not been reported in any other material system. Based on that we suggest that the origin may lie in the (near) degeneracy of the top of the valence band of Si, absent in e.g. GaAs 2DEGs and InAs nanowires. If a quantum dot is alternately filled by heavy and light holes which have different types of orbitals, the coupling of the wave functions to the leads can differ for consecutive hole numbers. This can induce strong variations in the capacitive coupling of successive orbitals and thus kites as we have observed. However, this does not identify \textsl{what} determines whether a stability diagram will consist of mainly parallelograms or kites, as in Figures 1a and 1b.

In these small diameter nanowires the degeneracy of the heavy hole and light hole subbands can be lifted by confinement, see e.g. calculations based on density functional theory \cite{dft1,dft2} and tight-binding models \cite{tb1,tb2}. If so, an empty Si nanowire quantum dot will start to fill with holes of the highest subband, resulting in a regular diamond pattern as in Supplementary Figure 1a. At higher charge numbers, holes of the second subband with a different effective mass, can also enter the dot and cause different slopes of adjacent diamond edges as in Supplementary Figure 1b. The latter device contains about 23-32 holes, whereas the first has about 3-13 holes. The higher hole number can thus cause the kites.

The results in Supplementary Figure 1 show that we can reproducibly fabricate Si nanowire quantum dots with sizes of 3 to 100 nm and a tuneable hole number over a large range. The observed asymmetry in Coulomb diamonds is a strong indication of the (near) degeneracy of heavy and light hole bands in silicon.\\

\noindent
\textbf{Quantization independent of magnetic field.}\\
\noindent
We have combined energy spectroscopy with magnetospectroscopy to study the magnetic field evolution of the discrete spectrum at the $N=2\leftrightarrow3$ and the $N=3\leftrightarrow4$ transition. Supplementary Figure 3a reveals a very pronounced grid of lines in the zoom of the $N=2\leftrightarrow3$ transition. The $N=2\leftrightarrow3$ transition is at a side gate voltage where the barriers are about equal in size, and as a result lines appear parallel to both diamond edges. Again we stress the contrast with the $N=0\leftrightarrow1$ and the $N=1\leftrightarrow2$ transition, where the barriers are asymmetric and the lines appear along one diamond edge. At the side gate voltage of the $N=3\leftrightarrow4$ transition one of the tunnel barriers becomes slightly more opaque, resulting in a lower differential conductance and less pronounced inelastic transport lines, see Supplementary Figure 3b.

Magnetospectroscopy of the energy levels is performed at side gate voltages close to both transitions, indicated by white dashed lines in Supplementary Figure 3a and 3b. With increasing magnetic field, the Coulomb blockaded region increases at the $N=2\leftrightarrow3$ transition (middle panel of Supplementary Figure 3a) and decreases at the $N=3\leftrightarrow4$ transition (middle panel of Supplementary Figure 3b). This is directly connected to the observed Coulomb peak evolution in Figure 4: the peak at the $N=2\leftrightarrow3$ transition moves \textsl{away }from the gate voltage at which the magnetospectroscopy is performed, whereas the peak at the $N=3\leftrightarrow4$ transition moves \textsl{towards} it. The lines above the ground-state transport level remain equidistant and again form a grid (red in the color scale). They neatly follow the ground-state transport line at a constant distance throughout the magnetic field sweep. The energy difference between the discrete levels is about 160--188 $\mu$eV at the $N=2\leftrightarrow3$ transition and 103--114 $\mu$eV at the $N=3\leftrightarrow4$ transition. We note that the Zeeman splitting of the $N=4$ diamond is faintly visible at the $N=3\leftrightarrow4$ transition.\\

\clearpage
\noindent
{\Large \textbf{Supplementary Figure captions}}\\

\noindent \textbf{Supplementary Figure 1: Quantum dots with larger hole numbers}
\noindent (\textbf{a}) Typical stability diagram, showing $\ln|dI/dV_{SD}|$ in color scale versus $V_{SD}$ and $V_{BG}$, revealing eleven Coulomb diamonds and a charge switch at $V_{BG}=-8.5$ V. (\textbf{b}) Stability diagram of the device in Figure 1b with kite-shaped diamonds. Grey and blue arrows indicate six pairs of excited-state lines parallel to the ground-state line of the next hole, with a slope different from the adjacent hole ground-state line (black and green arrows). Cotunnelling processes appear in red, e.g. in the two leftmost diamonds in (\textbf{a}), and in almost all diamonds in (\textbf{b}). We estimate the number of charges by counting up holes from zero, starting at the backgate voltage at which the dot is emptied at a high bias (100 mV). We combine this pinch-off voltage of $V_{BG}=-1.3$ V with $C_{BG}=0.2$ aF to estimate $N$ in (\textbf{a}) to be 7. Analogously, $C_{BG}=0.7$ aF and high-bias pinch-off at $V_{BG}=9$ V indicate that $M$ in (\textbf{b}) is about 29. (\textbf{c}) Capacitances from source, drain and gate to the dot of (\textbf{b}) versus number of holes, $M$. $C_{BG}$ is nearly constant because the backgate is relatively far away. $C_{S}$ and $C_{D}$ vary strongly from 2 to 10 aF. Ellipses (arrows) indicate pairs of source and drain capacitances that have comparable (different) values, as a result of equal (unequal) orbital coupling. Here we have assumed the $M^{\textrm{th}}$ diamond in (\textbf{b}) to be 29. The capacitances values at $M=29$ are calculated from the slopes at the transition from 28 to 29 holes, where the $29^{\textrm{th}}$ orbital is used for transport.\\

\noindent \textbf{Supplementary Figure 2: Zeeman energy in a few-hole Si nanowire quantum dot.}
\noindent (\textbf{a}) Stability diagram of a single-hole silicon nanowire quantum dot in another device. Here we need 30 V on the backgate to go from one to two holes as a result of a very small capacitance to the dot. The excited-state of the second hole is about 120 meV. Both the small capacitive coupling and the large level splitting indicate a very small quantum dot ($<10$ nm). This device exhibits switching behavior at the 0$\leftrightarrow$1 transition, probably due to bistable potential fluctuations caused by a charge in the environment of the dot. (\textbf{b}) Zoom of the 0$\leftrightarrow$1 transition at 0 and 9 T, revealing the Zeeman splitting at positive bias, cf. Figure 3a.\\

\noindent \textbf{Supplementary Figure 3: Quantization independent of magnetic field.}
\noindent (\textbf{a,b}) Zooms on the 2$\leftrightarrow$3 and the 3$\leftrightarrow$4 transitions at 0 T (leftmost panels) and 9 T (rightmost panels). Magnetospectroscopy of the discrete energy levels of the 2$\leftrightarrow$3 and the 3$\leftrightarrow$4 transition (middle panels), taken at the line cuts indicated by white dashed lines. The excitation lines above the ground-state transport line (yellow in color scale) remain equidistant and form a grid (red in color scale). They neatly follow the ground-state transport line at a constant distance throughout the magnetic field sweeps. A charge switch causes a discontinuity at $\textrm{B}=3.7$ T in (\textbf{a}). Blue dotted arrows indicate the Zeeman-split excited state.

\clearpage

\noindent
{\Large \textbf{Supplementary Figures}}\\

\begin{figure}[h]
\centering
\textbf{Supplementary Figure 1:\\ Quantum dots with larger hole numbers}\\
\vspace{1cm}
\includegraphics[width=.95\textwidth]{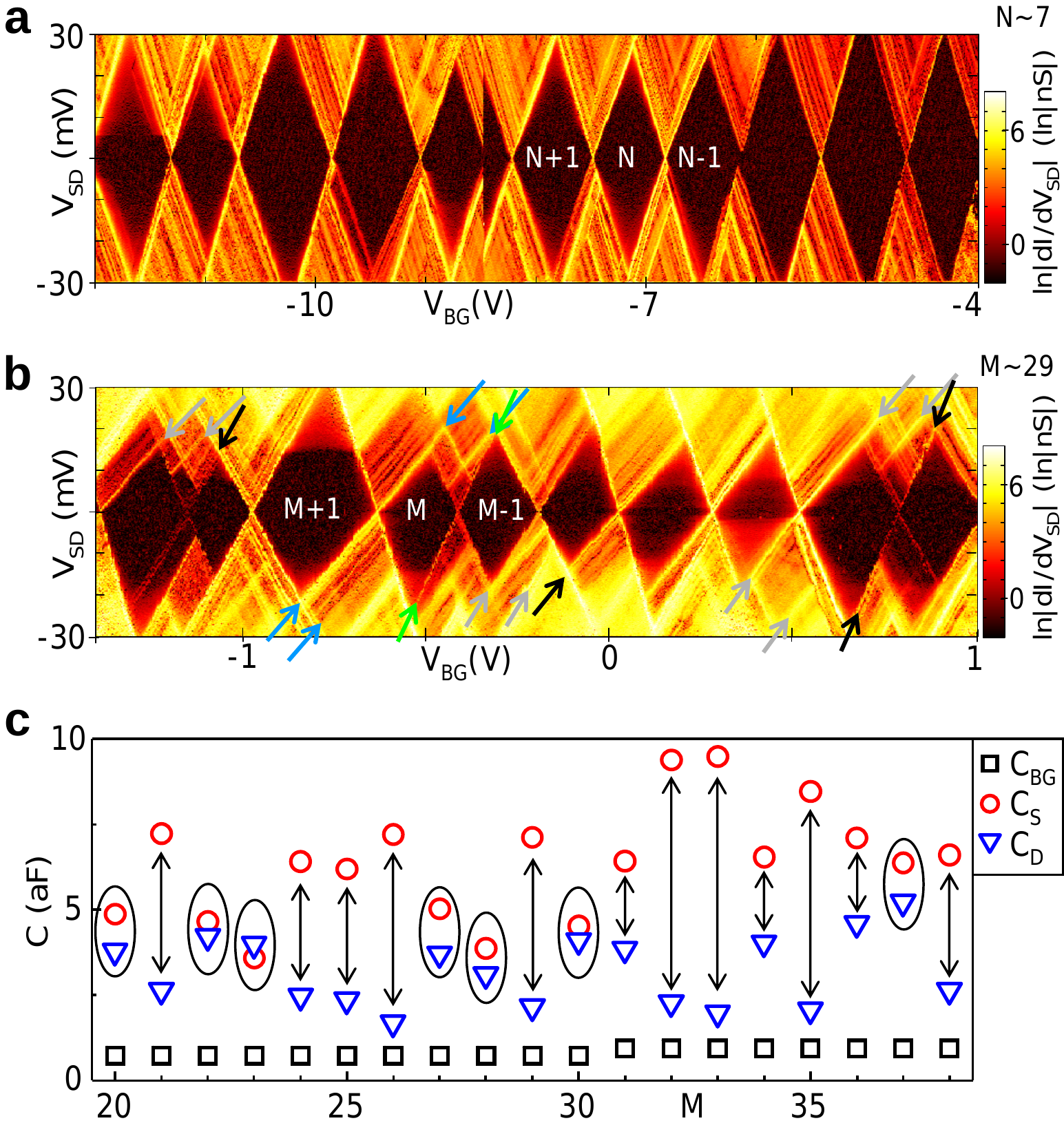}
\label{supplfig1}
\end{figure}

\clearpage
\begin{figure}[h]
\centering
\textbf{Supplementary Figure 2:\\ Zeeman energy in a few-hole Si nanowire quantum dot.}\\
\vspace{1cm}
\includegraphics[width=.9\textwidth]{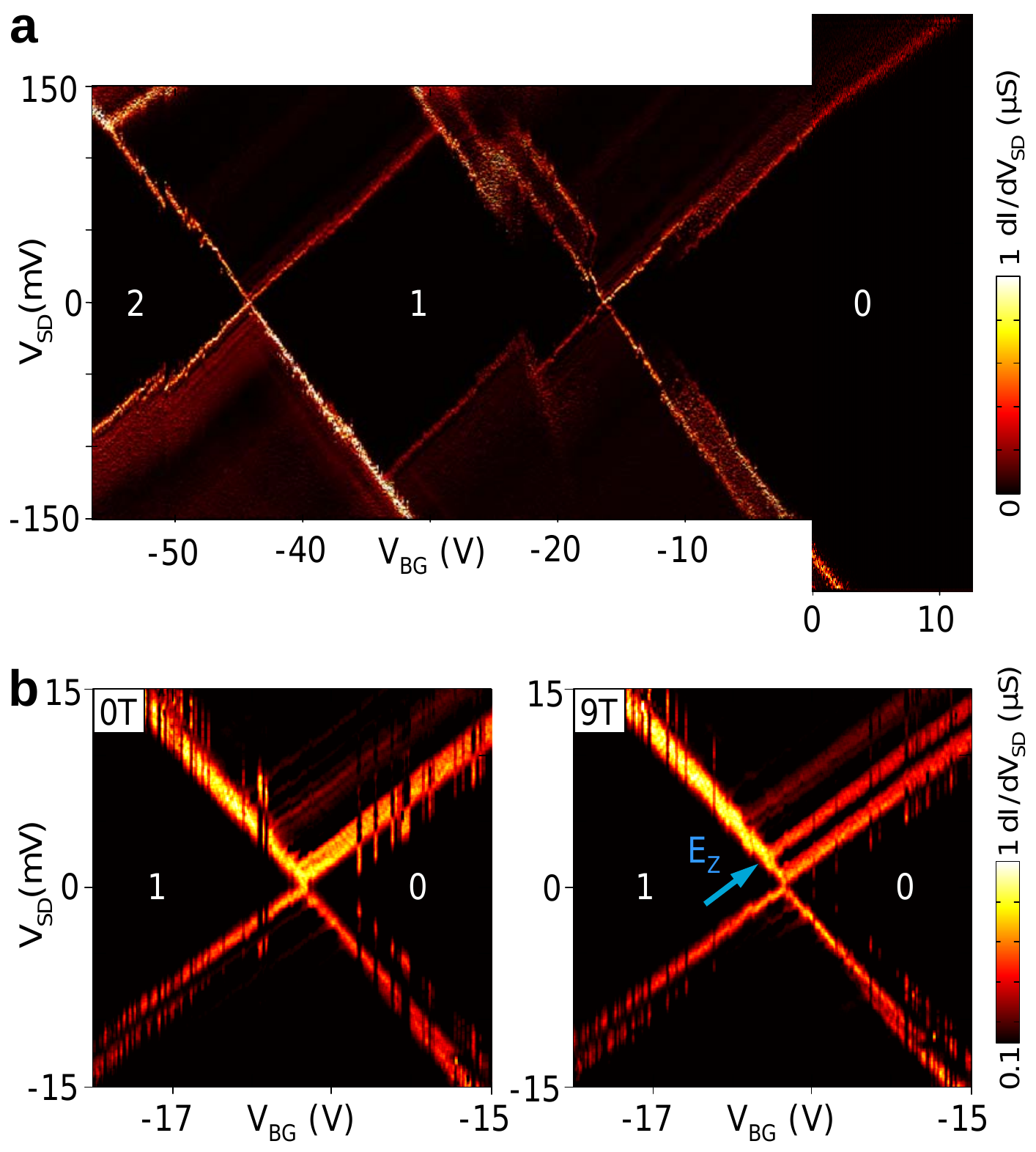}
\label{fhdfig6}
\end{figure}

\clearpage
\begin{figure}[h]
\centering
\textbf{Supplementary Figure 3:\\ Quantization independent of magnetic field.}\\
\vspace{1cm}
\includegraphics[width=\textwidth]{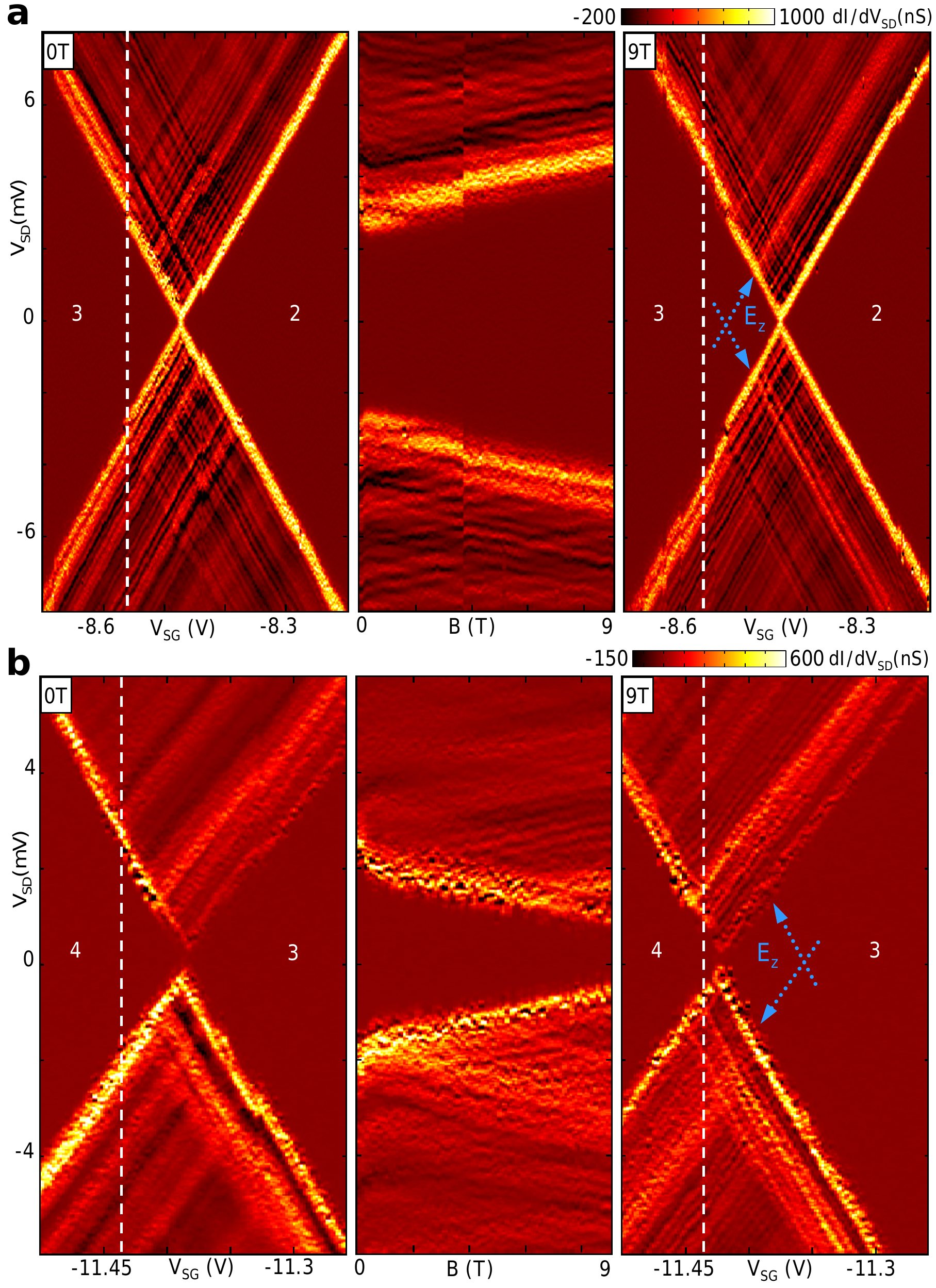}
\label{supplfig3}
\end{figure}

\end{document}